\documentclass[aps,twocolumn,prb,amsmath,amssymb,superscriptaddress,longbibliography]{revtex4-2}
\usepackage{graphicx}
\usepackage[colorlinks=true,urlcolor=blue,citecolor=blue,linkcolor=blue]{hyperref}
\usepackage{amsmath}
\usepackage{amssymb}
\usepackage{amsthm}
\usepackage{amsfonts}
\usepackage{enumerate}
\usepackage{subfigure}
\usepackage{centernot}
\usepackage{tikz}
\usepackage{ulem}

\newcommand{\affiliationUSST}{School of Physics, University of Shanghai for Science and Technology, Shanghai 200093, China}

\begin{document}

\title{Monopole Spin Density Wave States in Magnetic Weyl Semimetals} 

 \author{Xi Luo}
 \thanks{Correspondence to: xiluo@usst.edu.cn}
  \affiliation {\affiliationUSST}
 %
 \date{\today}

 \begin{abstract}
 The interplay between topology and magnetism in Weyl semimetals has recently emerged as a fertile ground for novel quantum phases. While monopole harmonic order parameters have been established for superconductivity and charge density waves in these systems, their spin density wave counterparts remain unexplored. Here we introduce monopole spin density wave (SDW) states arising from particle-hole pairing between nested Fermi surfaces enclosing Weyl nodes of the same chirality. We demonstrate that the SDW order parameter inherits a nontrivial pairing Berry phase and is described by monopole harmonic functions that exhibit topologically protected nodal structures in the gap function. Through a concrete lattice model, we show that helical and cycloidal SDW orders produce distinct signatures in band structures, Fermi arc distributions, and surface spin polarization patterns, which can be directly resolved by spin- and angle-resolved photoemission spectroscopy. Remarkably, we find that the quantum geometric tensor of the monopole pairing realizes ideal quantum geometry in the weak-coupling limit, where quantum distance fluctuations are entirely governed by Berry curvature. Our results not only unify the understanding of monopole ordered states across pairing channels but also provide experimental avenues for distinguishing competing magnetic orders in ReAlX (Re=rare earth elements, X=Si, Ge) materials and suggest potential applications in topological spintronics.
 \end{abstract}
\maketitle

{\bf  \textit{{Introduction}}.} {The discovery of topological semimetals has significantly expanded our understanding of topological states of matter \cite{Wan2011,Huang2015,Lv2015a}. In Weyl semimetals, the band-crossing points, Weyl nodes, act as sources and sinks of Berry curvature, and carry a definite topological charge (chirality). When doped such that Fermi surfaces (FSs) form around the type-I Weyl nodes, these topologically nontrivial FSs provide a unique platform for the emergence of novel ordered states.}

In recent years, it has been recognized that pairing between FSs in Weyl semimetals can inherit the nontrivial Berry phases of single-particle states, giving rise to a new class of ordered states, namely monopole harmonic ordered states \cite{nagaosa2003,Haldane2015,Garrido2020,Frazier2026,Li2020,Uchoa2026}. The seminal work by Li and Haldane revealed that when zero-momentum Cooper pairing occurs between two FSs carrying opposite Chern numbers, the Berry connection of the pairing operator is the difference of the two single-particle Berry connections, yielding a total Berry flux of $4\pi q_p$ \cite{Haldane2015}. This nontrivial "pairing Berry phase" prevents the pairing order parameter from being globally defined on the entire FS, thereby forcing the gap function to develop topologically protected nodal structures, whose total vorticity is uniquely determined by the pairing monopole charge $q_p$. Consequently, the pairing symmetry is no longer described by conventional spherical harmonics but by monopole harmonic functions $\mathcal{Y}_{q;l,m}$. This theoretical framework has spawned the study of novel topological ordered states such as monopole superconductivity (SC) \cite{Haldane2015} and monopole charge density waves (CDW) \cite{Li2020}. In the monopole CDW states, when electron-like and hole-like FSs enclosing Weyl nodes of the same chirality are nested, the particle-hole pairing similarly acquires a nontrivial Berry phase.  The gap function nodes emerge as zero-energy Weyl nodes, accompanied by nontrivial Fermi arc surface states \cite{Li2020}.

Recently, experimental observations of helical magnetic order in the ReAlX (Re=rare earth elements, X=Si/Ge) family of materials have provided important clues for the realization of monopole spin density wave (SDW) states \cite{White2020,Broholm2021,Philip2021,Tafti2023,Li2023,Hirschberger2025,Hirschberger2026}. SmAlSi possesses a nearly isotropic crystal electric field, allowing helical magnetic order to stabilize. The neutron diffraction experiments by Yao et al. confirmed an incommensurate helical magnetic order in SmAlSi with wave vector ${\bf Q}=({\frac{1}{3}-\delta,\frac{1}{3}-\delta,0})$, where $\delta=0.007$ \cite{Tafti2023}, precisely matching the nesting vector between Weyl Fermi pockets of the same chirality. More importantly, they identified an A-phase in the finite-temperature-finite-field region of the magnetic phase diagram and observed a topological Hall effect therein, suggestive of topological magnetic excitations such as skyrmions. These experimental results demonstrate that Weyl Fermions can directly participate in magnetic interactions via the Ruderman-Kittel-Kasuya-
Yosida mechanism \cite{Xiao2015,Askari2015,Zhou2015,Zhou2017,Liang2025}, driving helical magnetic order.

These advances naturally raise an important question: Can helical SDW order also exhibit monopole harmonic symmetry? In other words, when SDW pairing occurs between topologically nontrivial FSs, does the order parameter similarly inherit a nontrivial pairing Berry phase and become described by monopole harmonic functions? This question has not yet been systematically addressed.

In this work, we present a systematic study of monopole SDW states. We demonstrate that when particle-hole pairing occurs between FSs surrounding two Weyl nodes of the same chirality, the SDW order parameter inherits a nontrivial pairing Berry phase, with its angular distribution described by the monopole harmonic functions $\mathcal{Y}_{-1;1,m}$. We further find that the quantum geometric tensor of the SDW pairing can be directly expressed in terms of single-particle quantum geometric tensors. Remarkably, in the weak-coupling limit monopole SDW states, the inequality between the quantum metric and Berry curvature, $\text{tr}(g_{ij})\geq |\boldsymbol{\Omega}|$, becomes saturated \cite{Roy2014}. This implies that the monopole SDW state provides an exactly solvable instance of ideal quantum geometry where the quantum distance fluctuations are entirely determined by the Berry curvature, analogous to the lowest Landau level \cite{Devereaux2015,Yang2021,Liu2022,Wang2026}. Through a concrete lattice model, we demonstrate the distinct signatures of helical and cycloidal orders in band structures, Fermi arc distributions, and surface spin polarization patterns. These differences can be directly probed by spin- and angle-resolved photoemission spectroscopy (spin-ARPES) \cite{Sun2015,Lv2015,xu2016}, providing a feasible route to distinguish between helical and cycloidal orders that are otherwise difficult to differentiate experimentally \cite{Tafti2023}. Furthermore, the nontrivial pairing Berry phase in monopole SDW states may induce fractionalized Fermi arc states and nonlinear transport responses, opening new possibilities for topological spintronic devices.

{\bf \textit{Monopole SDW}}. The topological Berry phases of pairing operators of SC  and CDW in Weyl semimetals and their connections with monopole harmonics have been discussed in the pioneering works \cite{nagaosa2003,Haldane2015,Li2020}. Here we focus on the monopole SDW states in magnetic Weyl semimetals.

We start with two Weyl nodes ${\bf K}^{(n)}$ with $n=1,2$, and assume their FSs are well nested, which we label FS$^{(n)}_{\chi, \gamma}$ where $\chi=\pm 1$ is the chirality of the Weyl node and $\gamma=(e,h)$ labels the type of Fermi pocket at the FS. Then, after projecting on the low energy FS$^{(n)}_{\chi,\gamma}$, the creation operator is defined as \cite{Li2020}
\begin{equation}
\alpha^{\dagger(n)}_{\chi,\gamma}=\sum_{a=\uparrow,\downarrow}\xi_{\chi,\gamma,a}({\bf k})c^\dagger_{a}({\bf K}^{(n)}+{\bf k}),
\end{equation}
where $a$ stands for (pseudo)spin degrees of freedom. $c^\dagger$ is the Fermion creation operator. ${\bf k}$ is the momentum relative to the ${\bf K}^{(n)}$ which lies on the surface by  shifting the FS$^{(n)}_{\chi,\gamma}$ towards the origin by $-{\bf K}^{(n)}$. $\xi_{\chi,\gamma}$ is the normalized helical eigenstate of spinor satisfying $\chi({\boldsymbol{\sigma}\cdot {\bf \hat{k}})\xi_{\chi,\gamma}}=\text{sgn}(\gamma)\chi\xi_{\chi,\gamma}$, where ${\boldsymbol{\sigma}}$ are the Pauli matrices acting on the spin space, $\hat{{\bf k}}$ is the unit vector along the ${\bf k}$ direction, $\text{sgn}(\gamma)$ is $+1$ for the electron pocket and $-1$ for the hole pocket. The single-particle Berry connection associated with $\xi_{\chi,\gamma}$ is defined as ${\bf A}_{\chi,\gamma}({\bf k})=\sum_ai\xi_{\chi,\gamma}^*({\bf k})\nabla_{\bf k}\xi_{\chi,\gamma}({\bf k})$, and its Berry curvature behaves as a monopole in  momentum space. The total Berry flux through FS$_{\chi,\gamma}$ is
\begin{equation}
   \Phi_{\chi,\gamma} =\oint_{\text{FS}_{\chi,\gamma}}d{\bf k}\cdot\nabla_{\bf k}\times {\bf A}_{\chi,\gamma}({\bf k})=-\text{sgn}(\gamma)\chi 2\pi.  \label{eq2}
\end{equation}
And the Chern number associated with FS$_{\chi,\gamma}$ is defined as Ch$_{\chi,\gamma}=\Phi_{\chi,\gamma}/(2\pi)$. Due to the monopole structure of the Berry curvature, which results in a nonzero Chern number on the FS, $\xi_{\chi,\gamma}$ cannot be globally well-defined on the FS. It can be resolved either by introducing a singularity string \cite{Dirac1931} or by choosing (at least) two different gauge patches which overlap over the equator and are related through a gauge transformation \cite{Yang1975}. For instance, by denoting $u_{\bf k}=\cos(\theta_{\bf k}/2)$ and   $v_{\bf k}=\sin(\theta_{\bf k}/2)e^{i\phi_{\bf k}}$, then one can choose $\xi_{+1,e}=(u_{\bf k},v_{\bf k})^T$ and $\xi_{+1,h}=(-v^*_{\bf k},u^*_{\bf k})^T$. In this gauge patch, the south pole ($\theta_{\bf k}=\pi$) is singular. One can define another gauge patch $u'_{\bf k}=e^{-i\phi_{\bf k}}u_{\bf k}$ and $v'_{\bf k}=e^{-i\phi_{\bf k}}v_{\bf k}$, then $\xi'_{+1,\gamma}$ is singular at the north pole ($\theta_{\bf k}=0$) and the corresponding Berry connection are related by gauge transformations ${\bf A}_{+1,e}'={\bf A}_{+1,e}+\nabla\phi_{\bf k}$ and ${\bf A}_{+1,h}'={\bf A}_{+1,h}-\nabla\phi_{\bf k}$ on the equator ($\theta_{\bf k}=\pi/2$). The signs of the gauge transformations are opposite for electrons and holes  because their Chern numbers are opposite in sign.

To achieve the monopole SDW order, we consider the pairing in the particle-hole channel of the same chirality \cite{Li2020}. The mean field Hamiltonian for SDW pairing reads,
\begin{equation}
    H_{\bf M}=\sum_{{\bf p},a,b} c_a^\dagger ({\bf p+Q})({\bf M}({\bf Q})\cdot \boldsymbol{\sigma})c_b({\bf p})+h.c., \label{HM}
\end{equation}
where ${\bf Q}$ is the {nesting vector}, ${\bf M}=M\hat{\bf n}$ is proportional to the magnetization up to an interaction strength, and $\hat{\bf n}$ is the unit vector of spin polarization. $M=|{\bf M}|$ is also known as the gap function. In the weak-coupling regime $M<<|\mu|$ where $\mu$ is the {chemical potential}, one can project the pairing Hamiltonian onto the helical FSs around the paired Weyl nodes,
\begin{equation}
    \tilde{H}_M=\sum_{\bf k}\alpha^{\dagger (1)}_{+1,e}({\bf k})\tilde{{ M}}({\bf Q,k})\alpha_{+1,h}^{(2)}({\bf k})+h.c.,
\end{equation}
where the projected gap function is 
\begin{eqnarray}
    \tilde{M}({\bf Q,k})&=&\sum_{a,b}\xi_{+1,e,a}^\dagger({\bf k}) ({\bf M}({\bf Q})\cdot \boldsymbol{\sigma}_{ab})\xi_{+1,h,b}({\bf k})\nonumber\\
    &=&\sum_{m=-1}^{+1}C_mM_m\mathcal{Y}_{-1;1,m}(\hat{\bf k}).
\end{eqnarray}
$M_0=M_z$, $M_{\pm1}=(M_x\mp iM_y)/\sqrt{2}$, and $C_{m}=-\sqrt{8\pi}/\sqrt{3}$. $\mathcal{Y}_{q;l,m}$ are the monopole harmonics \cite{Haldane2015} with $q$, $l(\geq q)$, and $m$ being the monopole charge, the total angular momentum, and  the magnetic quantum number, respectively (see {Appendix} A). The emergence of monopole harmonics is due to the nontrivial Berry phase of the pairing operator. One can define the SDW pairing operator associated with the first term in $\tilde{H}_M$ as
\begin{equation}
    \hat{P}_{SDW}({\bf k})=\alpha^{\dagger (1)}_{+1,e}({\bf k})(\boldsymbol{\sigma}\cdot \hat{\bf n})\alpha_{+1,h}^{(2)}({\bf k}).
\end{equation}
Since $\boldsymbol{\sigma}\cdot \hat{\bf n}$ does not depend on ${\bf k}$, the pairing Berry phase can be obtained from the single-particle ones, i.e., ${\bf A}_{SDW}={\bf A}_{+1,e}^{(1)}-{\bf A}^{(2)}_{+1,h}$ \cite{nagaosa2003,Haldane2015,Li2020}. The monopole charge of $\hat{P}_{SDW}$ reads $\oint d{\bf k}\cdot \nabla_{\bf k}\times {\bf A}_{SDW}\equiv4\pi q=-4\pi$, which is consistent with the monopole harmonics used in the projected gap function  $\hat{M}$. In this notation, the monopole charge associated with ${\bf A}_{+1,\gamma}$ is $-\text{sgn}(\gamma)\frac{1}{2}$. 

Therefore, after projection, the low-energy effective two-band Hamiltonian reads
\begin{equation}
    H_{eff}=\Psi({\bf k})^\dagger
    \left(
    \begin{matrix}
        v_F|{\bf k}|-\mu & \tilde{M}({\bf Q,k}) \\
        \tilde{M}^*({\bf Q,k}) & -v_F|{\bf k}|+\mu
    \end{matrix}
    \right)\Psi({\bf k}),
\end{equation}
where $\Psi({\bf k})=(\alpha^{(1)}_{+1,e}({\bf k}),\alpha^{(2)}_{+1,h}({\bf k}))^T$, $v_F$ is the Fermi velocity. The gap function $\tilde{M}$  exhibits nodes due to the nature of the monopole harmonics. For instance, a case similar to a magnetic order with nonzero $M_0$ only has been discussed in \cite{Li2020} (which is interpreted as a {CDW} in the pseudospin space). In this case, the nodes are identified as Weyl nodes by directly checking the behavior of the gap function $\tilde{M}$ and the effective Hamiltonian near the north pole and  south pole for a fixed $k_z$. These nodes are Weyl nodes with the same chirality as $\alpha_{\chi,\gamma}^{(n)}$ which do not depend on the details of the mechanism of SDW ordering. The emergence of low-energy Weyl Fermions is inherited from the topology at high energy and is topologically protected \cite{nagaosa2003,Haldane2015,Li2020}. 

{\bf \textit{Lattice model}.} To further investigate the properties of monopole SDW, we consider a lattice model of a doped Weyl semimetal with nested FSs,
\begin{eqnarray}
    H&=&\sum_{{\bf p}}c^\dagger({\bf p})h_0({\bf p})c({\bf p})+H_M, \label{htotal}\\
    h_0&=&t_x\sin p_x\sigma_x\tau_x+t_y\sin{p_y}\sigma_y\tau_x+t_z\sin p_z\tau_y\nonumber\\
    &&+(m+2-\cos k_x-\cos k_y-m\cos 2k_z)\tau_z\nonumber\\
    &&+m_z\sigma_z+m_x\tau_x+V_0\cos k_z.
\end{eqnarray}
$t_x$, $t_y$, and $t_z$ are hopping constants which are set to unity. ${\boldsymbol{\tau}}$ is the Pauli matrix that acts on the orbital space. $m_z$ and $m_x$ terms break time-reversal and inversion symmetries, respectively. The Weyl nodes of $h_0$ are located along the $p_z$ axis and are determined by
\begin{equation}
    4m^2 \sin^4p_z+\sin^2p_z+m_x^2=m_z^2.
\end{equation}
Without loss of generality, we choose $m=2$, $m_z=1.5$, and $m_x=1$. Then, there are four Weyl nodes ${\bf{P}}^{(1)}=(0,0,-\frac{5\pi}{6})$, ${\bf{P}}^{(2)}=(0,0,-\frac{\pi}{6})$, ${\bf{P}}^{(3)}=-{\bf{P}}^{(2)}$, and ${\bf{P}}^{(4)}=-{\bf{P}}^{(1)}$. The chiralities of ${\bf{P}}^{(1)}$ and ${\bf{P}}^{(3)}$ are $-1$ and those of ${\bf{P}}^{(2)}$ and ${\bf{P}}^{(4)}$ are $+1$. $V_0$ term (effective chemical potential) controls the size of FSs of the Weyl nodes and dose not alter their locations.

$H_M$ is the mean field SDW pairing Hamiltonian described by Eq. (\ref{HM}). In general, the SDW pairing can stem from the Coulomb interaction. However, there is another origin of SDW ording in magnetic Weyl semimetals without inversion symmetry, such as the ReAlX materials {\cite{Tafti2023}}. In such systems, non-collinear magnetic order, for instance, helical and cycloidal magnetic orders are favored by finite {Dzyaloshinskii-Moriya} interaction. 
The Kondo coupling between magnetic order and local spin can generate the SDW order. These magnetic orders have probably been observed in SmAlSi with a nearly nesting momentum that connects two Weyl nodes of the same chirality, but they are hardly distinguishable from each other by {neutron diffraction measurements \cite{Tafti2023}}. Here we consider two kinds of magnetic orders, the helical order ${\bf M}_1=(\sin Qz,\cos Qz, 0)$ and the cycloidal order ${\bf M}_2=(0,\sin Qz, \cos Qz)$. The SDW pairing can be achieved by the Kondo coupling $\sum_i\kappa_1{\bf M}_1\cdot {\bf s}_i$ and $\sum_i\kappa_2{\bf M}_2\cdot {\bf s}_i$ where $\kappa_1$ and $\kappa_2$ are the Kondo coupling strengths, and ${\bf s}_i=c_i^\dagger(\frac{\boldsymbol{\sigma}}{2})c_i$ is the local spin at the lattice site $i$. The kondo interaction has the same form as the mean field Hamiltonian (\ref{HM}) of the SDW pairing. For the monopole SDW, it should pair between Weyl nodes with the same chirality in the particle-hole channel. Therefore, in our lattice model, we choose $Q=\pi$ (nesting between ${\bf P}^{(2)}$ and ${\bf P}^{(4)}$). Then, diagonalizing the total Hamiltonian (\ref{htotal}) reduces to solving the Harper equation \cite{Kohmoto1989,Kohmoto1990,Hatsugai1993,Luo2025}. And in the momentum space, the helical and cycloidal orders correspond to ${\bf M}_1({\bf Q})=(0,1,0)$ and ${\bf M}_2({\bf Q})=(0,0,1)$, respectively (see Appendix B).

\begin{figure*}	\includegraphics[width=0.8\textwidth]{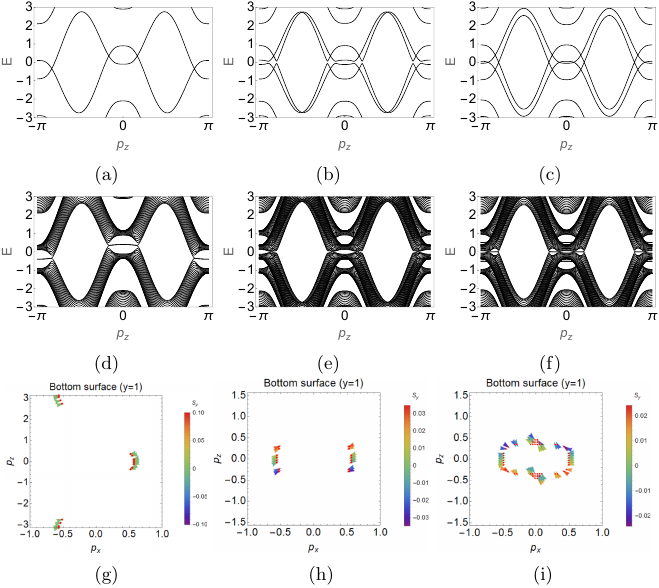}
\caption{(color online) (a), (b), and (c) are the bulk bands along $p_z$ axis. $p_x=p_y=0$ and $V_0=0.4$. (a) Without the $H_M$ term. (b) $\kappa_1=0.4$ and $\kappa_2=0$. (c) $\kappa_1=0$ and $\kappa_2=0.4$. (d), (e), and (f) are the band structures with finite length ($N_y=20$ sites) along $y$ direction that correspond to (a), (b), and (c), respectively. (g), (h), and (i) are the spin polarizations along the Fermi arcs of the bottom surfaces in (d), (e), and (f), respectively. (e) and (f) are plotted within the folded (magnetic) Brillouin zone. The red points stand for the momentum locations of the Fermi arc states. The arrows indicate the directions of spins in the $S_x$-$S_z$ plane and their color labels the magnitude of $S_y$. }
\label{fig1}	
\end{figure*}

We plot the band structures and spin polarizations of the Fermi arcs in Fig. \ref{fig1}. With $h_0$ only, there are Fermi arcs connecting ${\bf P}^{(1)}$ to ${\bf P}^{(4)}$ and ${\bf P}^{(2)}$ to ${\bf P}^{(3)}$ as expected (see Fig. \ref{fig1}d). For the helical order ${\bf M}_1$,  the corresponding projected gap function $\tilde{\bf M}_1$ in the momentum space is proportional to $\mathcal{Y}_{-1;1,1}-\mathcal{Y}_{-1;1,-1}$ whose nodes are shifted away from the north pole and the south pole. Therefore, a finite gap opens near zero energy in the $p_z$ axis (see Fig. \ref{fig1}b). For the cycloidal order ${\bf M}_2$,  the projected gap function $\tilde{\bf M}_2$ is proportional to  $\mathcal{Y}_{-1;1,0}$ which has nodes at the north pole and the south pole  (see Fig. \ref{fig1}c). This case is equivalent to the example of CDW discussed in Ref. \cite{Li2020}, and our results are consistent with those in Ref. \cite{Li2020}. For a strong interaction (large $\kappa$), there is a Lifshitz transition: the Weyl nodes collide and open a gap, and the Weyl semimetal turns into an insulator.

Moreover, the spin polarizations along the Fermi arcs show different patterns. When the SDW order is absent, the locations of the Fermi arcs are asymmetric in ${p_x}$ direction in the bottom surface (see Fig. \ref{fig1}g). The spin component  $S_y$  is zero along the Fermi arcs. For the helical order, the Fermi arc states near the boundary $p_z=\pm \pi$ of Fig. \ref{fig1}g are folded to $p_z=0$ by the nesting vector ${\bf Q=(0,0,\pi)}$. Unlike the free case, the $S_y$ component is nonzero and changes signs near the two ends of a Fermi arc (see Fig. \ref{fig1}h). This sign-changing behavior indicates the monopole SDW pairing has a nontrivial Berry phase. For the cycloidal order, there are extra Fermi arc states near $p_x=0$ besides the Fermi arc states inherited from the free Weyl semimetal (see  Fig. \ref{fig1}i). In both cases, nonzero $S_y$ is generated by the non-collinear magnetic order. Although the helical and cycloidal orders are difficult to distinguish from each other in ReAlX by neutron diffraction measurements \cite{Tafti2023}, it would be possible to determine the type of magnetic order by spin-ARPES \cite{Sun2015,Lv2015,xu2016} if monopole SDW is realized in ReAlX materials, since these two orders generate different band structures, spin polarizations, and Fermi arc distributions for the monopole SDW states.  

As a comparison, we provide an example of SDW pairing in the particle-hole channel of Weyl nodes of opposite chiralities in Appendix C. We find that the Fermi arc distributions are asymmetric with respect to $p_x$ axis in the reduced Brillouin zone on one surface, and the $S_y$ does not change signs at the two ends of the Fermi arc which indicates the Berry phase of the SDW pairing is trivial.

{\bf \textit{Pairing quantum geometry and its topological constraints of the monopole SDW state.}} The quantum geometric tensor for a state $|\psi\rangle$ is defined as 
\begin{equation}
    Q_{ij}=\langle \partial_i\psi|\partial_j\psi\rangle-\langle\partial_i\psi|\psi\rangle \langle\psi| \partial_j\psi\rangle=g_{ij}-\frac{i}{2}\Omega_{ij},
\end{equation}
where $\Omega_{ij}$ is the Berry curvature and $g_{ij}$ is known as the quantum metric \cite{Vallee1980,Berry1989}. Since the pairing Berry phase can be expressed in terms of the Berry phases of single Weyl nodes, it is natural to ask whether the pairing quantum geometric tensor is related to the single-particle ones. The answer is yes. For simplicity, let us consider the monopole CDW pairing operator $\hat{P}_{CDW}=\alpha^{\dagger}_{+1,e}\alpha^{}_{+1,h}$ \cite{Li2020}. After a direct calculation, the pairing quantum geometric tensor reads
\begin{eqnarray}
    Q_{ij}^{P}&=&g_{ij}^P-\frac{i}{2}\Omega_{ij}^P,\\
g_{ij}^{P}&=&g_{ij}^{+1,e}+g_{ij}^{+1,h}=\frac{1}{2|{\bf k}|^2}(\delta_{ij}-\hat{k}_i\hat{k_j}),\\
    \Omega_{ij}^P&=&\Omega_{ij}^{+1,e}-\Omega_{ij}^{+1,h}=-\epsilon_{ijm}\frac{\hat{k}_m}{|{\bf k}|^2}.
\end{eqnarray}
This result also works for the monopole SDW pairing state and monopole superconductivity \cite{Haldane2015} in Weyl semimetals. Furthermore, there is an inequality between the quantum metric and Berry curvature, $\text{tr} (g_{ij})\geq |\boldsymbol{\Omega}|$ where $|\boldsymbol{\Omega}|$ is the norm of the Berry curvature \cite{Roy2014,Fu2024}. This inequality saturates when the Weyl nodes are isotropic (in the weak coupling limit of the monopole CDW/SDW pairing). One can directly verify 
\begin{equation}
    \text{tr} (g_{ij}^P)=\frac{1}{|{\bf k}|^2}=|\boldsymbol{\Omega}^P|.
\end{equation}
This provides an example of a gapless system that saturates this inequality. When the inequality is saturated, the quantum distance fluctuations of the wave function are entirely determined by the Berry curvature, which is known as the ideal quantum geometry \cite{Devereaux2015,Yang2021,Liu2022,Wang2026}. Ideal quantum geometry has been studied for stabilizing fractional quantum Hall states in the lowest Landau level and fractional Chern insulator states \cite{Wang2026}. Besides the ideal quantum geometry case, in general, the quantum geometry also contributes to nonlinear transport in non-centrosymmetric systems, such as the nonlinear Hall effect and nonlinear thermal transport \cite{Fu2015,Mak2019,Ma2019,Yin2021,Yan2025,xu2025,Uchoa2025,Si2025}. The stiffness of flat band superconductivity/superfluid also originates from the quantum geometry \cite{Peotta2015,Liang2017,Huhtinen2022,Yang2025}.  A more detailed study of the nonlinear Hall effect and potential fractionally charged Fermi arc states arising from the quantum geometry of monopole CDW/SDW states in magnetic Weyl semimetals will be presented in future work.

{\bf \textit{Conclusions.}} In this work, we have systematically investigated monopole SDW states in magnetic Weyl semimetals. 

When particle-hole pairing occurs between nested FSs surrounding Weyl nodes of the same chirality, the SDW order parameter inherits a nontrivial pairing Berry phase. The projected gap function is naturally expressed in terms of monopole harmonic functions $\mathcal{Y}_{-1,1,m}$, with the monopole charge $q=-1$ determined by the topological structure of the underlying FSs. This provides a unified description of monopole ordered states across the particle-particle (superconductivity) and particle-hole (CDW and SDW) channels.

Through a concrete lattice model realizing the monopole SDW state, we revealed distinct experimental signatures for helical and cycloidal magnetic orders. These two types of non-collinear orders exhibit markedly different behaviors in bulk band structures, Fermi arc distributions, and surface spin polarization patterns. These provide a concrete route to resolve the long-standing ambiguity between helical and cycloidal orders in ReAlX materials, where neutron diffraction alone has proven insufficient.

We uncover a deep connection between monopole SDW states and ideal quantum geometry. The pairing quantum geometric tensor can be directly constructed from single-particle quantum geometric tensors, and in the isotropic Weyl-node limit, the inequality $\text{tr} (g_{ij})\geq |\boldsymbol{\Omega}|$ becomes saturated. This establishes the monopole SDW state as an exact gapless realization of ideal quantum geometry, akin to the lowest Landau level, where quantum distance fluctuations are entirely determined by Berry curvature. This finding enriches the landscape of ideal quantum geometric systems beyond fractional Chern insulators. 


Looking forward, several interesting directions merit further investigation. The competition and coexistence between monopole SDW, monopole CDW, and monopole SC orders in the same material system, particularly in ReAlX compounds where multiple pairing instabilities may be proximate, remain  an open question. The nonlinear transport signatures arising from the ideal quantum geometry of monopole SDW states, including the nonlinear Hall effect and nonlinear thermal transport, warrant detailed theoretical and experimental study. Finally, the potential realization of fractionally charged surface states due to the nontrivial pairing Berry phase in monopole SDW systems presents an exciting frontier for exploring unconventional quantum matter. These phenomena open new avenues for topological spintronic devices that exploit the interplay between band topology, quantum geometry, and non-collinear magnetism.


\textit{Acknowledgements.} We thank Yue Yu and Long Liang for helpful discussions. We would also like to thank Congjun Wu for his excellent introductory talk on monopole harmonics. This work is supported by the National Natural
Science Foundation of China with Grants No.~12174067. 

\textit{Data availability.} The data that support the findings of this article are not publicly available. The data are available from the authors upon reasonable request.



\newpage

\appendix

\begin{widetext}

\section{Detailed calculations of $\tilde{M}$} \label{app1}
In the main text, the projected gap function reads,
\begin{equation}
    \tilde{M}({\bf Q,k})=\sum_{a,b}\xi_{+1,e,a}^\dagger({\bf k}) ({\bf M}({\bf Q})\cdot \boldsymbol{\sigma}_{ab})\xi_{+1,h,b}({\bf k}),
\end{equation}
where $\xi_{+1,e}=(\cos(\theta_{\bf k}/2),\sin(\theta_{\bf k}/2)e^{i\phi_{\bf k}})^T$ and $\xi_{+1,h}=(-\sin(\theta_{\bf k}/2)e^{-i\phi_{\bf k}},\cos(\theta_{\bf k}/2))^T$. Then one has the following results,
\begin{eqnarray}
    \xi_{+1,e}^\dagger \sigma_z\xi_{+1,h}&=&-\sin(\theta_{\bf k})e^{i\phi_{\bf k}},\\
    \xi_{+1,e}^\dagger \frac{\sigma_x+i\sigma_y}{\sqrt{2}}\xi_{+1,h}&=&-\sqrt{2}\cos^2(\frac{\theta_{\bf k}}{2}),\\
    \xi_{+1,e}^\dagger \frac{\sigma_x-i\sigma_y}{\sqrt{2}}\xi_{+1,h}&=&-\sqrt{2}\sin^2(\frac{\theta_{\bf k}}{2})e^{-2i\phi_{\bf k}}.
\end{eqnarray}
By comparing with the standard monopole harmonics \cite{Haldane2015},
\begin{eqnarray}
    \mathcal{Y}_{-1;1,0}&=&\sqrt{\frac{3}{8\pi}}\sin(\theta)e^{i\phi}, \\
    \mathcal{Y}_{-1;1,+1}&=&\sqrt{\frac{3}{8\pi}}\cos^2(\frac{\theta}{2}), \\
    \mathcal{Y}_{-1;1,-1}&=&\sqrt{\frac{3}{8\pi}}\sin^2(\frac{\theta}{2})e^{-2i\phi}, 
\end{eqnarray}
one obtain
\begin{eqnarray}
    \tilde{M}({\bf Q,k})=-\sqrt{\frac{8\pi}{3}}\sum_{m=-1}^{+1}M_m\mathcal{Y}_{-1;1,m}(\hat{\bf k}),
\end{eqnarray}
with $M_0=M_z$, $M_{\pm}=(M_x\mp iM_y)/\sqrt{2}$.

\section{Harper equation for the lattice model} \label{app2}

The mean field Hamiltonian for the helical SDW order ${\bf M}_1=(\sin Qz, \cos Qz,0)$ in the real space reads
\begin{equation}
    H_{M_1}=\sum_i\kappa_1 c_i^\dagger ({\bf M}_1\cdot \frac{\boldsymbol{\sigma}}{2})c_i,
\end{equation}
where $\kappa$ is the Kondo coupling strength and $i$ is the lattice site. Define the partial Fourier transformation along the $z$ direction,
\begin{eqnarray}
    c_{z}=\int\frac{dp_z}{2\pi}c_{p_z} e^{ip_zz},
\end{eqnarray}
then,
\begin{equation}
    H_{M_1}=\sum_z\int\frac{dp_z}{2\pi}\int \frac{dp'_z}{2\pi}(c^\dagger_{\uparrow,p_z}e^{-ip_zz},c^\dagger_{\downarrow,p_z}e^{-ip_zz})\left(
    \begin{matrix}
        0 & -\kappa_1ie^{iQz}/2\\
      \kappa_1  ie^{-iQz}/2 & 0
    \end{matrix}
    \right)
    \left(
    \begin{matrix}
        c_{\uparrow,p'_z}e^{ip'_zz} \\
        c_{\downarrow,p'_z}e^{ip'_zz}
    \end{matrix}
    \right),
\end{equation}
where ${p_x}$ and $p_y$ are omitted. Clearly, each ${p_z}$ is coupled to ${p_z+Q}$ and ${p_z-Q}$. For the monopole SDW order of the lattice model considered in the main text, we choose $Q=\pi$. Then in the basis $(C_{\uparrow,p_z+\pi},C_{\downarrow,p_z+\pi},C_{\uparrow,p_z},C_{\downarrow,p_z})^T$, $H_{M_1}$ corresponds to
\begin{equation}
\frac{\kappa_1}{2}
\left(
    \begin{matrix}
    0 & 0 & 0 & -i\\
    0 & 0 & i & 0\\
    0 & -i  & 0 &0\\
    i & 0 & 0 &0\\
    \end{matrix}
    \right).
\end{equation}
By comparing with Eq. (3) in the main text, one finds that ${\bf M}_1({\bf Q})=(0,1,0)$. And The diagonalization of the total lattice model is reduces to solving the Harper equation \cite{Kohmoto1989,Kohmoto1990,Hatsugai1993,Luo2025}. Similarly, one can obtain that the cycloidal magnetic order ${\bf M}_2=(0,\sin \pi z, \cos \pi z)$ corresponds to ${\bf M}_2({\bf Q})=(0,0,1)$.

\section{SDW pairing in the particle-hole channel of Weyl nodes with opposite chiralities} \label{app3}
To compare with the monopole SDW states in the magnetic Weyl semimetal, we consider the SDW states with $Q=2\pi/3$ for the lattice model used in the main text, which is a nesting vector between two Weyl nodes with opposite chiralities. Use the same method described in Appendix B, in the basis $(C_{\uparrow,p_z+Q},C_{\downarrow,p_z+Q},C_{\uparrow,p_z+2Q},C_{\downarrow,p_z+2Q},C_{\uparrow,p_z},C_{\downarrow,p_z})^T$, the helical magnetic order $H_{M_1}$ and the cycloidal magnetic order $H_{M_2}$ correspond to 
\begin{equation}
    \left(
    \begin{matrix}
    0 & 0 & 0 & 0 &0 & -i\\
    0 & 0 & i & 0 & 0 & 0\\
    0 & -i  & 0 &0 &0 & 0\\
    0 & 0 & 0 & 0 & i & 0\\
    0 & 0 & 0 & -i & 0 &0\\
    i & 0 & 0 & 0 & 0 & 0
    \end{matrix}
    \right), \quad
     \left(
    \begin{matrix}
    0 & 0 & 1/2 & 1/2 &1/2 & -1/2\\
    0 & 0 & -1/2 & -1/2 & 1/2 & -1/2\\
    1/2 & -1/2  & 0 &0 &1/2 & 1/2\\
    1/2 & -1/2 & 0 & 0 & -1/2 & -1/2\\
    1/2 & 1/2 & 1/2 & -1/2 & 0 &0\\
    -1/2 & -1/2 & 1/2 & -1/2 & 0 & 0
    \end{matrix}
    \right),
\end{equation}
respectively. We plot the band spectra and spin polarizations of the Fermi arcs in Fig. \ref{supp1}. By compareing with Fig. 1 in the main text, we find that the SDW with the nesting between Weyl nodes with opposite chiralities behaves completely different from the monopole SDW states. The spectrum of helical magnetic order is gapless while it is gapped for cycloidal order (see Fig. \ref{supp1}a and Fig. \ref{supp1}d). The Fermi arc states exist over the entire  Brillouin zone (see Fig. \ref{supp1}b and Fig. \ref{supp1}e). On the bottom surface, the Fermi arc is asymmetric with respect to $p_x=0$. Although the magnitude of $S_y$ is small, is does not flip signs at the two ends of the Fermi arc, which indicates the pairing Berry phase is trivial.

\begin{figure*}	\includegraphics[width=0.8\textwidth]{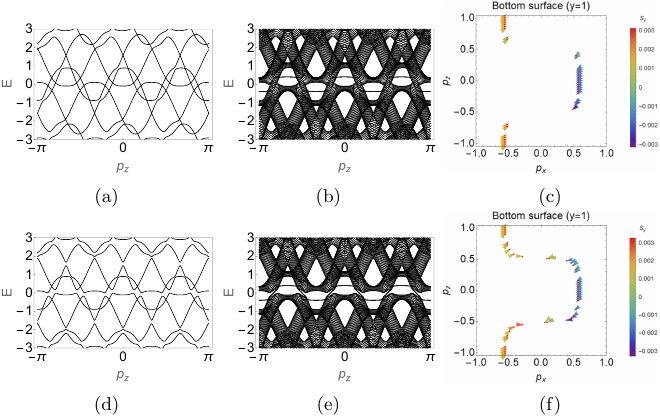}
\caption{(color online) (a) and (d) are the bulk bands along $p_z$ axis for $Q=2\pi/3$. $p_x=p_y=0$ and $V_0=0.4$.  (a) $\kappa_1=0.4$ and $\kappa_2=0$. (d) $\kappa_1=0$ and $\kappa_2=0.4$. (b) and (e) are the band structures with finite length ($N_y=20$ sites) along $y$ direction that correspond to (a) and (d), respectively. (c) and (f) are the spin polarizations along the Fermi arcs  within the folded Brillouin zone of the bottom surfaces in (b) and (e), respectively. The red points stand for the momentum locations of the Fermi arc states. The arrows indicate the directions of spins in the $S_x$-$S_z$ plane and their color labels the magnitude of $S_y$. }
\label{supp1}	
\end{figure*}

\end{widetext}

\end{document}